 \documentclass[final,5p,times,twocolumn,authoryear]{elsarticle}

\usepackage{amssymb}
\usepackage{graphicx}
\usepackage{dblfloatfix}
\usepackage{subcaption}
\usepackage{multirow}
\usepackage[ruled]{algorithm2e} 

\usepackage{float}
\usepackage{booktabs}
\usepackage{makecell}  
\usepackage{tabularx}
\usepackage[T1]{fontenc}
\usepackage{booktabs} 
\usepackage{enumitem}
\usepackage{breakurl}

\makeatletter
\def\ps@pprintTitle{%
  \let\@oddhead\@empty
  \let\@evenhead\@empty
  \let\@oddfoot\@empty
  \let\@evenfoot\@oddfoot
}
\makeatother

\begin{document}

\begin{frontmatter}


\author{Baris Yamansavascilar\corref{cor1}\fnref{label1}}
\ead{baris.yamansavascilar@boun.edu.tr}
\author{Atay Ozgovde\fnref{label1}}
\ead{ozgovde@boun.edu.tr}
\author{Cem Ersoy\fnref{label1}}
\ead{ersoy@boun.edu.tr}


\cortext[cor1]{Corresponding author.}


\title{Dynamic Capacity Enhancement using Air Computing: An Earthquake Case}

\address[label1]{Department of Computer Engineering, Bogazici University, Istanbul, Turkey}



\begin{abstract}

Earthquakes are one of the most destructive natural disasters harming life and the infrastructure of cities. After an earthquake, functioning communication and computational capacity are crucial for rescue teams and healthcare of victims. Therefore, an earthquake can be investigated for dynamic capacity enhancement in which additional resources are deployed since the surviving portion of the infrastructure may not meet the demand of the users. In this study, we propose a new computation paradigm, air computing, which is the air vehicle assisted next generation edge computing through different air platforms, in order to enhance the capacity of the areas affected by an earthquake. To this end, we put forward a novel paradigm that presents a dynamic, responsive, and high-resolution computation environment by explaining its corresponding components, air layers, and essential advantages. Moreover, we focus on the unmanned aerial vehicle (UAV) deployment problem and apply three different methods including the emergency method, the load balancing method, and the location selection index (LSI) method in which we take the delay requirements of applications into account. To test and compare their performance in terms of the task success rate, we developed an earthquake scenario in which three towns are affected with different severity. The experimental results showed that each method can be beneficial considering the circumstances, and goal of the rescue.



\end{abstract}




\end{frontmatter}


\section{Introduction}
\label{intro}





On February 6, 2023, two powerful earthquakes with magnitude 7.8 and 7.5 hit south and central Turkey (\cite{guo2023preliminary}). They caused widespread damage in the region regarding human lives and infrastructure as the earthquake epicenter was close to crowded cities including Gaziantep and Kahramanmaras (\cite{world2023turkiye}). Since the damage was so heavy, more than 140000 people from 94 countries joined the corresponding rescue efforts after Turkey's call for international help. Currently, the confirmed death toll is above 50000 people (\cite{deathToll}). 




One of the most important problems in the affected region was the deprivation of communication. Since most of the infrastructure had collapsed, computers, base stations and communication links could not serve properly to people including the rescue teams, media, and earthquake victims who desperately needed to organize properly. Therefore, this disaster has shown once more that a dynamic capacity enhancement scheme for the continuity of communication and computation services regarding extraordinary events must be considered.

To meet the dynamic requirements of the next-generation computer networks and different application types that have diverse service level agreements (SLA), vertical networking opportunities are recently used through low altitude platforms (LAP), high altitude platforms (HAP), and low earth orbit (LEO) satellites (\cite{guo2021service, liu2018space, baltaci2021survey}). Each of these air platforms can provide different opportunities in terms of latency, data rate, computational capability, coverage, and mobility to the corresponding applications using the benefits of 3D networking. However, the joint operation of those air platforms in a dynamic environment to meet different requirements have not been properly applied and therefore still be investigated.

\begin{figure}[t]
\centering
\includegraphics[scale=0.045]{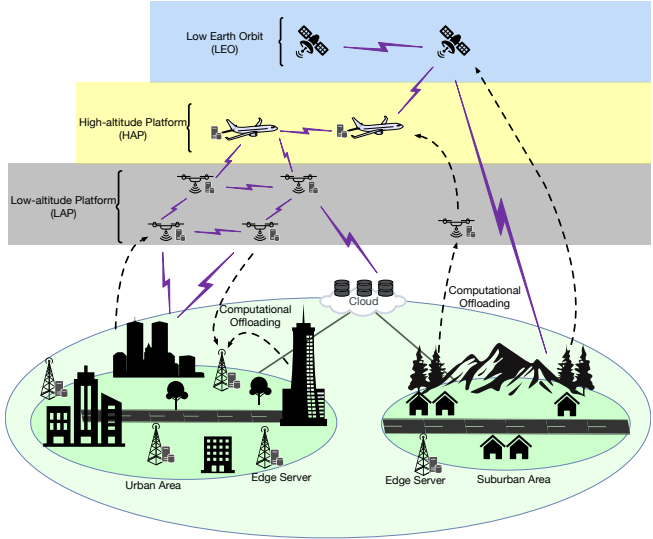}
\DeclareGraphicsExtensions.
\caption{Air Computing Architecture.}
\label{AirComputing}
\end{figure}

Considering the task offloading paradigms such as fog and edge computing, air vehicles are also used for enhancing the computational capacity. Currently, the most popular implementation of this approach is the deployment of unmanned aerial vehicles (UAVs) as UAV-assisted mobile edge computing ( \cite{li2018uav}). However, even though other air vehicles including airplanes, balloons, and low earth orbits (LEOs) can also be used for this purpose, they are deployed as standalone units. Since the collaborative utilization of those air units can open new horizons for efficient resource allocation, task offloading, and content caching, we believe that edge computing would evolve into this 3D paradigm. To this end, we consider the organization and execution of these air platforms regarding the computational needs of the applications as a new computation paradigm. Thus, we propose the name of \textit{air computing} considering air vehicle assisted next generation edge computing. The architecture of air computing is depicted in Figure \ref{AirComputing}.

The locations where the existing infrastructure would be insufficient for the particular application requirements can use air computing for task offloading, content caching, and resource allocation. Hence, there are many advantages that air computing can provide using different air vehicles for several real-world scenarios including remote health, real-time video, augmented reality, outdoor activities, and natural disasters (\cite{wang2018power, zeng2016wireless}). Especially, this practice is crucial for an area that has faced a disaster such as an earthquake. 




In this study, we focus on the dynamic capacity enhancement using UAVs for the computational needs of applications that may not be executed satisfactorily because of the overloaded terrestrial servers in the disaster area. The cause of the overloading can be a result of either the destruction of the existing terrestrial servers or the increasing number of tasks in the area due to the panic induced by the disaster, computational requirements, and increased need for communication. Thus, the organization and deployment policies of UAVs are crucial to provide the required quality of service (QoS). 

We introduce three different methods for the deployment of UAVs that can be used in the case of a disaster. To this end, we present a simulation for an earthquake scenario that affects three towns differently. In the simulation, there are several events that alter the existing infrastructure, interarrival times of tasks, and number of users. Thus, we compare the methods based on their reaction to those events by considering the success rate of user applications which may have different SLA requirements. Our main contributions in this study can be summarized as follows:

\begin{itemize}

\item We propose a new paradigm called air computing, which is the next generation edge computing through organized air vehicles such as UAVs, airplanes, and LEOs.

\item We focus on the problem of UAV deployment under the management of a HAP vehicle for enhancing the computational capabilities in the case of a disaster, which is not thoroughly investigated in the literature.

\item We show how different UAV deployment methods behave in the case of an earthquake. To highlight the significance of this case, we experiment with different events that can cause fluctuations in load and capacity in three different towns. 

\item We elaborate three heuristic methods that can be used for UAV deployment, and compare their performance with each other. Note that each of these methods provides different benefits based on the goals and condition of the post-disaster area.


\end{itemize}


The rest of this paper is organized as follows. In Section 2, we elaborate the related works including task offloading, edge computing, UAVs. We introduce the air computing paradigm in Section 3. In Section 4, we  and describe the system model including UAV deployment methods. Section 5 presents the earthquake scenario, our simulation metrics, and the performance evaluation of our system along with the experiments. Finally, we provide conclusions in Section 6. 

\section{Related Work}


Task offloading has been widely studied in edge and cloud computing (\cite{senyo2018cloud}). Especially, delay-intolerant application tasks can be offloaded to the edge (\cite{feng2022computation, laroui2021edge}). In (\cite{peng2021constrained}), authors focused on meeting QoS considering varied constraints of IoT devices in an edge environment. To this end, they proposed three constrained multi-objective evolutionary algorithms (CMOEAs) to solve time and energy consumption problems for offloading scenarios in edge computing. Feng et al. studied mission-critical IoT services by considering the optimization of green task offloading and priority-differentiated queuing policies in (\cite{feng2021energy}). They developed a priority-differentiated offloading strategy by taking QoS requirements of mission-critical applications into account. Moreover, they benefited the Lyapunov optimization technique considering the energy consumption (\cite{neely2010stochastic}). In (\cite{xue2021joint}), authors focused on the task offloading and resource allocation. They proposed a dynamic incentive mechanism for a multi-user and multi-vehicle vehicular edge computing environment. Moreover, they used the Stackelberg game (\cite{zhang2009stackelberg}) for the interaction between users and service providers. Xu et al. investigated efficient data routing paths regarding each offloaded task in (\cite{xu2021throughput}). Their objective was to maximize the network throughput in terms of the number of successfully completed tasks. They formulated the problem into a Integer Linear Programming (ILP) problem subject to several constraints including the task delay and network resource capability. In (\cite{chen2021energy}), authors studied on an energy efficient offloading strategy. To this end, they developed a self-adaptive particle swarm optimization algorithm, which reduces the system energy consumption.

Since UAVs can be deployed dynamically for networking and task offloading, the load on the terrestrial resources can be alleviated (\cite{huda2022survey}). Therefore, required QoS for the corresponding applications can be met otherwise it would be difficult to provide. In (\cite{seid2021multi}), the authors proposed a multi-agent deep reinforcement learning (MADRL) based method to minimize the computation costs in terms of energy consumption and computation delay. To test the performance of their method, they developed a multi-UAV enabled IoT edge environment using a single controller  Software-Defined Networking (SDN). Apostolopoulos et al. proposed a task offloading system in which partial offloading of the tasks can be carried out using edge servers and UAVs in (\cite{apostolopoulos2021data}). Hence, they formulated a maximization problem using the principles of Prospect Theory (\cite{kahneman2013prospect}) by considering the optimal user task offloading to the available computing choices. In (\cite{el2021uav}), the authors focused on mission-critical applications considering UAV-aided ultra-reliable low-latency computation offloading. Consequently, they divided the problem into two phases. In the first phase, they considered the optimization of the UAV placement problem. In the second phase, they took the offloading and resource allocation decisions into account for QoS. Thus, they formulated these issues as non-convex mixed-integer programs. Wang et al. proposed a two-layer optimization method considering deployment of UAVs and task scheduling in (\cite{wang2019joint}). They also considered  offloading decisions and resource allocation with the purpose of minimizing the system energy consumption. In (\cite{zhan2021multi}), the authors proposed a framework for a multi-UAV-enabled edge system regarding the task offloading and resource allocation. Their goal was to maximize the number of served IoT devices. They formulated the corresponding optimization problem as a mixed integer nonlinear programming (MINLP). Afterwards, they developed an iterative algorithm to solve the problem.

Recently, studies has focused on the disaster cases using UAVs. In (\cite{wang2022task}), the authors addressed the battery and computational resource limitation of UAVs in a fog computing environment. To this end, they proposed a practical task offloading scheme for UAVs, which use ground resources. Note that the tasks were offloaded from UAVs in their study. For this purpose, they developed a stable matching algorithm to match each UAV with a ground resource. Shah et al. investigated the maximization of an utility function considering a disaster scenario in UAV-assisted edge computing networks in (\cite{shah2023mobile}). They proposed an algorithm which jointly optimizes computational capacity, UAVs location, user association, required delay, and coverage. They compared the performance of their method with a random and greedy schemes. In (\cite{jin2023equalizing}), the authors addressed the problem of the fairness of ground users served by UAVs in a post-disaster rescue. Therefore, they jointly optimized the number of service grids, flight trajectory, and hovering position of UAVs. Thus, they proposed a fair service policy model by using an adaptive area division and consolidation method. Kaleem et al. proposed a reinforcement learning based greedy algorithm to meet required QoS and minimum rate of users in (\cite{kaleem2022enhanced}). They also addressed that an unplanned UAV deployment can cause an interference from the neighboring co-channel stations. Their simulation results showed that their proposed scheme outperformed the conventional water filling algorithm. In (\cite{wang2023secure}), the authors addressed the security threats on UAVs during the data transmissions for UAV-assisted disaster rescue. Hence, they developed a secure information sharing scheme by considering task offloading from UAVs to ground vehicles. In (\cite{do2021joint}), the authors focused on a joint optimization of resource allocation and real-time deployment regarding a UAV-based relay system in emergency cases. They considered UAVs as flying relay nodes for the communication. Therefore, they proposed a new k-means clustering model with several QoS constraints to optimize UAV deployment. They also aimed to maximize energy efficiency in terms of overall transmitted data of users in the disaster area.

\subsection{Differences between existing works}
To the best of our knowledge, studies that work on aerial unit support in the literature do not consider the dynamic events in a disaster scenario. We believe that realistic handling of dynamic events are essential for the performance evaluation of a system since the reactive response of the corresponding methods is crucial in an emergency situation. Moreover, related studies also do not consider different application requirements which is also important for the overall system performance. In this study, we focus on how the corresponding UAV deployment methods react based on different events such as destruction of the existing resources, and increased number of users which affects the load. Furthermore, we evaluate the conditions of different towns so that we are able show that taking the application requirements into account is essential for an acceptable system performance. 

\section{Air Computing}



The essential idea of edge computing is that offloading the computation-intensive tasks from end devices to the corresponding edge servers since battery and CPU limitations cannot allow the local execution. Therefore, users decide where to offload and when to offload the corresponding tasks if there are multiple edge servers nearby. However, since the delay requirements change for the mission-critical applications, and mobile devices including tablets and smartwatches proliferate, traditional edge computing based on terrestrial resources would be insufficient to meet the suitable computing capacity. As a remedy, air based computational resources are recently proposed to enhance the computational capacity by augmenting 3D networking opportunities.

The most popular implementation of this paradigm is UAV-assisted edge computing  since UAVs provide flexibility in terms of flying, and lower latency since they are close to the ground. However, other air vehicles including airplanes, balloons, and LEOs are also used for this purpose. Since there is no conventional title for this concept considering the organization of these air vehicles regarding computational task offloading, we propose the name of air computing. Thus, air computing includes all air vehicles in order to enhance the edge computing paradigm. As a result, we believe that air computing is the evolution of  edge computing through air vehicles.

\subsection{Air Platforms}
Air computing consists of three air platforms including LAP, HAP, and LEO. Each platform provides different features regarding the requirements of the underlying environment. 

\subsubsection{LAP}

The main deployment of LAP is on urban areas in which the existing infrastructure is built well and therefore meets the QoS of user applications. Since the operational altitude of the corresponding air vehicles in LAP, which are UAVs, is below 10 km, the propagation delay would change between 10 - 30 $\mu$s. Moreover, since they can provide Line of Sight (LoS), connectivity, service provision, and latency can be ensured seamlessly.

Another important feature of LAP is the flexible utilization of UAVs considering their easily configurable stationary positions. Therefore, UAVs can use their computational capacity efficiently since they can easily move to areas where user density is high. Note that apart from their computational assistance for air computing, they can also be used as relay nodes in order to provide seamless mobility.

Even though LAP provides many benefits in dynamic environments, energy consumption is an important issue for UAVs since they need charging stations for their batteries. Moreover, this maintenance should be performed daily as their batteries are limited. Note that the energy consumption of UAVs can be affected by weather conditions and rains, especially when they move horizontally against the wind.

\subsubsection{HAP}

Air vehicles in HAP can be used on urban and suburban areas since they can fly at high altitudes between 10 - 30 km. Therefore, their propagation delay changes between 50 - 85 $\mu$s. Moreover, channel and weather conditions also affect  communication quality. Thus, they cannot be used for mission-critical applications whose delay tolerance is low.

The essential use case for HAP is for regional coverage in which airplanes and balloons can be deployed as management nodes for UAVs and terrestrial servers. Moreover, they can be used as computational resources if the SLA requirements of the corresponding tasks would not be violated by the delay. Note that even though the delay is higher in HAP, it would still be lower than the cloud server option which is deployed in the wide area network (WAN). On the other hand, even though they may provide a relatively stationary position, they cannot use their capacity as efficiently as UAVs since their configurability is not flexible.

One of the most important advantages of balloons and airplanes is that they can fly for days since they use fuel as their energy source. Moreover, the effect of weather conditions would be limited because of their altitude. Thanks to these important features, HAP can be used for long-distance communication, energy consumption, and management opportunities.

\subsubsection{LEO}

LEO platform consists of satellites whose altitude changes between 160 - 2000 km. Because of this altitude range, its propagation delay would be between 1.5-3 ms which is not suitable for low latency applications. However, they can carry out edge computing solutions through task offloading as either used as a relay node or using their limited onboard capacity. Moreover, they are also used to access cloud computing solutions.

Based on their features, LEO is generally utilized for rural areas whose access to computing resources is limited. Because of this reason and their high speeds, their capacity would be wasted. On the other hand, they can give service for months since they can meet their required energy from sonar power. Moreover, their energy consumption cannot be affected by weather conditions due to their altitude.

\subsection{Advantages}

Air computing provides many advantages regarding task offloading, content caching, latency, coverage, and mobility. Considering task offloading, which is the essential use-case of air computing, 3D networking opportunities employ various computational technologies through different air layers. For example, if the existing infrastructure is built well such as an urban area, air computing would consist of only edge servers or UAV-assisted edge servers. On the other hand, if there is limited infrastructure in a suburban or rural area, UAVs, airplanes, and LEOs would be organized to meet the required QoS. Thus, task offloading can be carried out without considering geographical disadvantages. Moreover, dynamic conditions of different environments would not affect the given service since air computing can adapt itself to the ever-changing environment. Furthermore, task offloading can be performed for both ground users and the users in the air. Hence, while users in the air can offload their tasks to the ground servers, terrestrial users can carry out the offloading using air vehicles.

Content caching is also another important execution of air computing. Since air vehicles can be used dynamically based on the needs, the corresponding contents related to a specific area or user groups can be cached easily. Therefore, users can access the requested pages, tools, and applications with low latency. Considering the fact that content caching optimizes three objectives including QoS guarantee, content popularity, and utility maximization (\cite{ouyang2018follow, zhang2018hierarchical}), users can benefit from the high hit ratio, which is the essential performance measurement for the quality of content caching optimization.

The coverage and therefore mobility are also performed with low latency through air computing. Since air computing can provide seamless coverage and pervasive connectivity through different air platforms, it can provide high mobility over 1000 km/hr. Moreover, the collaboration of different air vehicles allows a smooth handover process of the corresponding tasks.


\section{System Model}

\begin{figure}[t]
\centering
\includegraphics[scale=0.11]{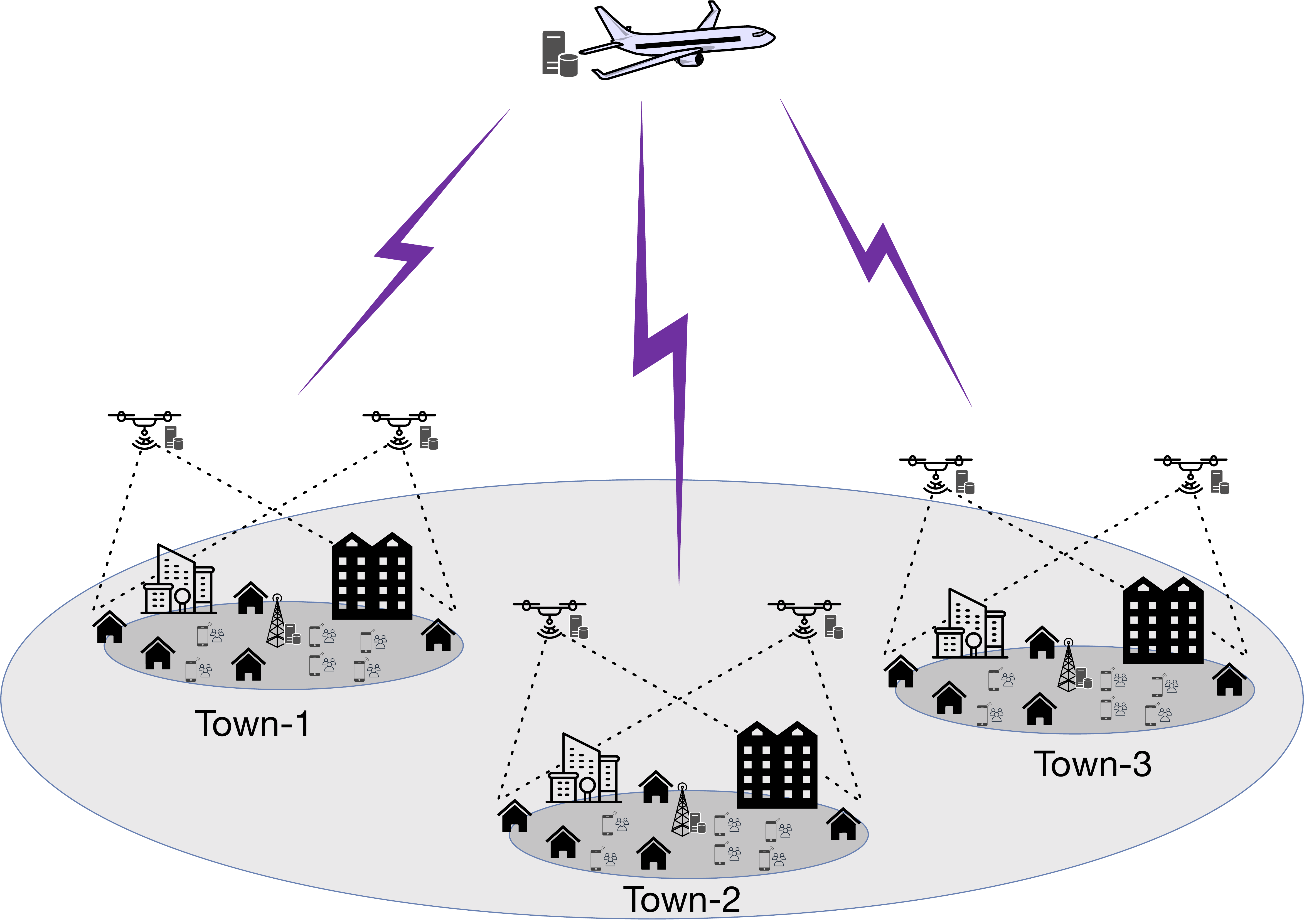}
\DeclareGraphicsExtensions.
\caption{Air computing environment in this study.}
\label{AirComputingEnvironment}
\end{figure}

In our air computing model, we design the environment considering the ground, LAP, and HAP layers as shown in Figure \ref{AirComputingEnvironment}. Each layer has significant components to carry out the air computing paradigm for dynamic capacity enhancement. The ground layer includes users, applications, tasks, towns, and edge servers. On the other hand, the LAP layer consists of UAVs which can fly dynamically to a corresponding location to enhance the capacity. The HAP layer comprises a corresponding air vehicle, which is an airplane, that covers each town. It is used as a management node for UAVs since it receives the most recent information about the environment. We detail each component in this section.

\subsection{User}

A user is a fundamental element for the ground layer in an air computing environment since their quality of experience (QoE) is the essential measurement for the performance of the system. A user can reside in an urban area, suburban area, or rural area to perform their tasks by utilizing air components. However, the demand for those air components may depend on the existing infrastructure and corresponding load. For example, if a user can exploit the edge servers in a Local Area Network (LAN), they may not need to use air components. However, on the other hand, if the load is extremely high with respect to the existing capacity in the network, then users would utilize air components to meet the expected QoS for their tasks. Note that user can run multiple applications considering different SLA requirements.

\subsection{Applications}

An application consists of atomic tasks each of which has a SLA requirement. If the latency which is required by SLA is exceeded, the corresponding task is assumed to be failed. Note that there would be multiple application types that have different latency requirements.

A task of an application can run on a user's device or it can be offloaded to either a UAV or an edge server to be processed. Hence, we are able to calculate the overall success rate of an application based on the success rate of its tasks.

\subsection{Edge Servers}

The primary computational resource in air computing is edge servers. An edge server can be placed in urban or suburban areas in which there is an infrastructure. They are preferred by applications to offload their task in order to meet  the corresponding QoS since edge servers provide lower network delay than cloud servers.

On the other hand, the capacity of an edge server is not as high as a cloud server. Therefore, if a task requires a high computational capacity and its delay tolerance is not low, it would be wise to be offloaded to a cloud server. We assume that each task in the environment has low delay tolerance and should be offloaded to one of the available  edge servers or UAVs. 

\subsection{UAVs}

We consider a UAV as a flying edge server with less capacity. They have similar features as edge servers however, since they are not connected to a ground power supply, and they are smaller, their computational capacity is lower than edge servers. Thus, they are used as the secondary computational resources to increase the capacity if needed in air computing. On the other hand, if there is no existing ground infrastructure, then they become the primary computational resource. 

The most important difference of a UAV regarding an edge server is that they are not always available physically for a particular location since they may have to relocate or recharge. Therefore, considering the offloading case, a UAV does not accept an offloaded task to process when it starts to relocate to another place.  

\subsection{HAP Vehicle}

Each UAV in our system is able to fly from one location to another based on the UAV deployment method computed and provided by the HAP vehicle throughout the environment. Since the airplane can cover the area including each town, it can receive the most recent information about users and UAVs through a separate communication channel. Thus, the connection information of users to the corresponding edge servers and UAVs, number of tasks, and the real-time location of UAVs can be provided to the airplane. Note that the airplane is not used for task offloading in this study; it is only utilized for the corresponding calculations, and management.

\subsection{Task Offloading}


Vertical networking opportunities through air computing provide important advantages for task offloading. In our system, since WAN delays prohibit delay intolerant tasks to be offloaded to the cloud, a user task can be offloaded to either an edge server or a UAV based on their availability.

Note that since our focus in this study is UAV deployment methods regarding the dynamic capacity enhancement for a disaster situation, we apply only a single policy for task offloading. Hence the offloading policy is that the corresponding task is offloaded to a server, including edge or UAV, whose estimated response time is the smallest. There would be many alternative methods for task offloading between an edge and UAV as their capacities are different, however we apply this method considering the disaster scenario. 

We assume that the bandwidth in LANs is very high so we presume that LAN delays are not load dependent. Thus, computational delays are the most decisive parameters for the success rate of tasks.


\subsection{Objective}

Our main goal for this study is to maximize the success rate of tasks. Each produced and offloaded task has a worst-case latency requirement based on the type of the corresponding application. Thus, a UAV deployment method applied to the active UAVs in the network is crucial for the performance of the environment.

\subsection{UAV Deployment Methods}

In order to demonstrate the dynamic capacity enhancement with different objectives, we propose three different methods for UAV deployment considering an environment where several towns may face a disaster. The disaster may be a flood, fire, hurricane, or earthquake. We manage each method on the HAP layer by collecting the corresponding statistical information including the number of tasks, user connectivity to edge servers, and UAV locations. Afterwards, we run the related methods on UAVs.

\subsubsection{Load Balancing Method}

Each city or town in an environment can be affected by the disaster differently. Therefore, the load produced by the applications can vary based on the different events such as rescue operations, increasing number of users, and panic. To this end, the load balancing method takes the number of tasks of each town offloaded in a predefined time period into account, and send a UAV to the location where the task count is the highest.

Considering the multiple UAVs and multiple towns, this method initially sends a UAV to the corresponding location and then subtracts a predefined number of tasks from its original number of tasks. Thus, the next UAV would be sent to another place if the original number of tasks of the towns is close to each other. Otherwise, if there is a large imbalance in the task density in those locations, the same location would be selected for the next UAV again. As a result, this method provides load balancing between different towns based on their task count. 

\subsubsection{Emergency Method}

The emergency method considers an area where the existing infrastructure is destroyed by the disaster. Therefore, it checks the users who are not able to offload their tasks to any edge server and UAV. After the detection of those users, a corresponding UAV is sent to the center location of those users.

In order to define the center location, this method uses the k-means algorithm by taking the location information of the corresponding users into account. Therefore, the value of $k$ depends on the environmental dynamics. Each destructed infrastructure of a town would increase the value of $k$ by one. Note that the number of UAVs in the environment should be higher than the value of $k$ to apply this method correctly. 

\subsubsection{Location Selection Index}

The Location Selection Index (LSI) method computes how many UAVs are required for an affected area considering the delay requirements of the application types. To this end, this method reactively takes the corresponding decisions.

Since the most important performance metric for a successfully completed task is its overall latency, the underlying infrastructure including edge and UAV should provide the necessary computation capacity. Moreover, that capacity should be considered for the existing load, which may change dynamically, in the area. To this end, the average delay in the corresponding area considering the capacity of UAVs and edge servers is calculated using an $M$/$M$/1 queueing model. Note that other models could also be used easily as long as they consider the current load on the system.

After the calculation of the average delay, our method compares it with the required delay considering the application types in the area. There are two essential cases that we should send UAV to increase the capacity: (1) If the required delay is smaller than the calculated average delay, (2) if the capacity of the resources, including edge servers and UAVs, is smaller than the load. Finally, based on the required capacity and capacity of UAVs, we compute the required number of UAVs to send the corresponding area.

\subsubsection{Random Method}

As a baseline, we also implemented the random method which randomly assigns each UAV to one of towns in the environment.

\section{Performance Evaluation}

We developed a simulation environment in order to test the performance of UAV deployment methods considering an earthquake scenario. We evaluated their performance based on the overall task success rate and the task success rate of the corresponding towns affected by the earthquake.

\subsection{Scenario}

\begin{figure}[t]
\centering
\includegraphics[scale=0.13]{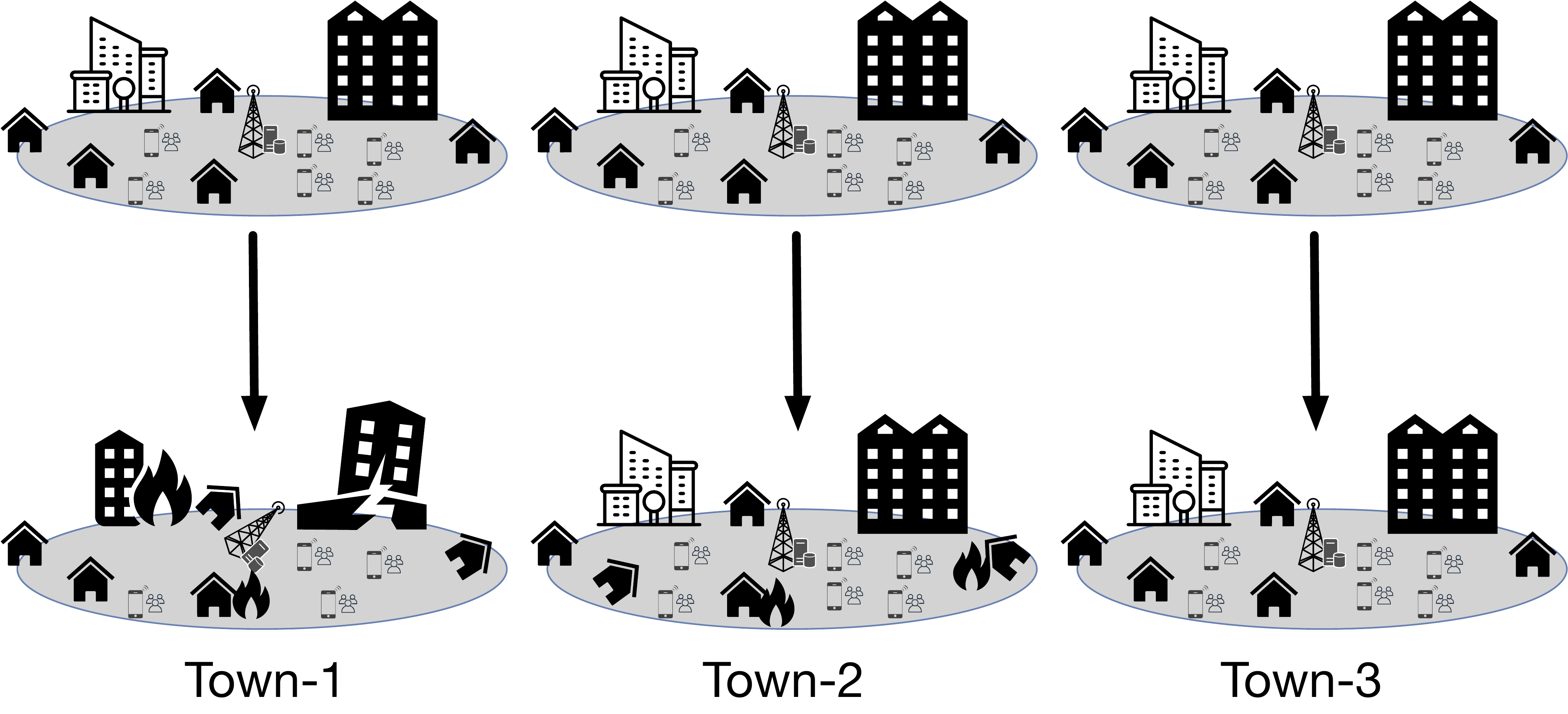}
\DeclareGraphicsExtensions.
\caption{The towns are affected by the earthquake differently in our scenario. Town-1 including its infrastructure is destroyed, while Town-2 is seriously affected. On the other hand, Town-3 has no damage but senses the disaster.}
\label{AffectedTowns}
\end{figure}

In our scenario, we have three towns which are affected by the earthquake at different extents. Town-1 is the most affected area as its infrastructure is completely destroyed. On the other hand, Town-2 severely senses the earthquake with no physical damage. Finally, Town-3 senses the earthquake mildly. The outcomes are depicted in Figure \ref{AffectedTowns}.

Each town has an edge server that is normally accessible by each user in the town. Initially, the computational capacity of each town is sufficient to meet the required worst case delay for the tasks of its users. However, after the earthquake, the demand of tasks, and the number of users change based on different events such as search and rescue activities, and rushing users. Therefore, additional capacity should be considered.

The duration of our simulation is 4000 seconds in which the earthquake happens at 1000 seconds. Before the earthquake, each user in Town-1 and Town-2 uses an application whose required number of CPU units, worst-case delay, and interarrival time are 90 units, 1-second, and 3.33 seconds, respectively. These parameters are the same for Town-3 except for the worst-case delay which is 2 seconds. 

After the earthquake at 1000 seconds, utilization of applications would differ based on the conditions of the towns and the effect on users. Therefore, the utilization of applications triples for each town due to the panic of people so that their new interarrival time is 1 second. Note that the edge server in Town-1 has been destroyed and out of service when the earthquake hit. 

Starting from 2000 seconds, the number of users doubles for each city because of several reasons. For Town-1, there are many rescue operations as the town is severely damaged. On the other hand, those operations are managed from Town-2 as its infrastructure has not been affected. Finally, aftershocks are observed using a facility in Town-3. Note that since rescue operations and their management are critical, the worst-case delay and interarrival time of the new users' tasks are 1 second with 90 CPU units. However, since the observation of aftershocks would not be as critical as those rescue operations, the worst-case delay for new users' tasks in Town-3 is 5 seconds with 12 CPU units. Moreover, their interarrival time is 1 second as aftershocks are frequent. 

\begin{figure}[t]
\centering
\includegraphics[scale=0.63]{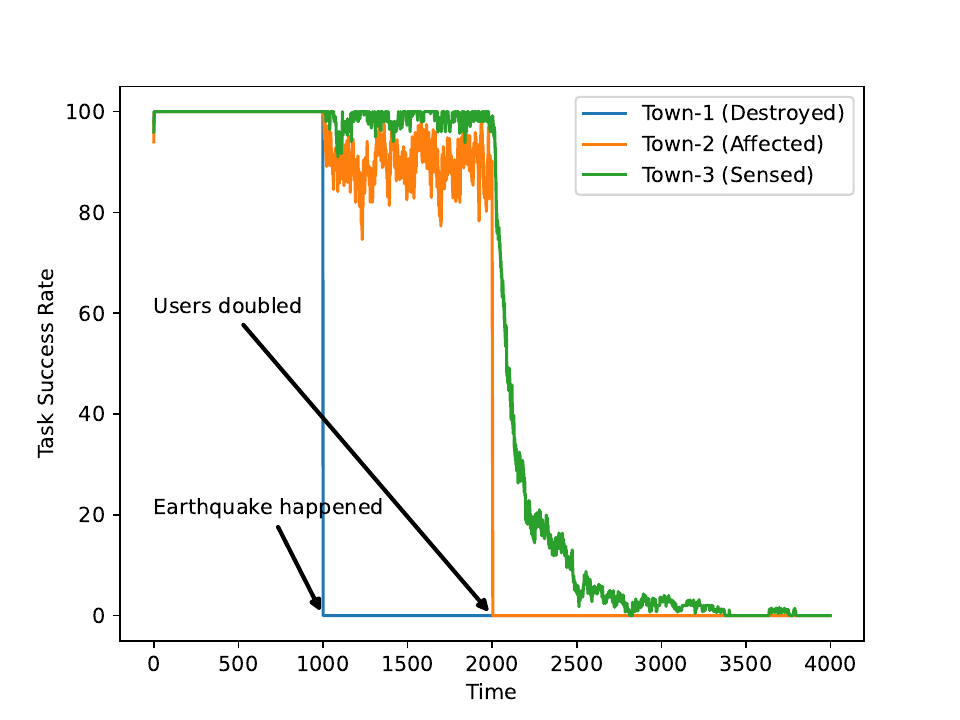}
\DeclareGraphicsExtensions.
\caption{The effect of the non-UAV policy in time domain. The earthquake happens at 1000 second, and the number of users in each town doubled at 2000 seconds.}
\label{NoUAVTime}
\end{figure}

\begin{figure*}[!t]

\begin{subfigure}{0.32\textwidth}
\includegraphics[width=\linewidth]{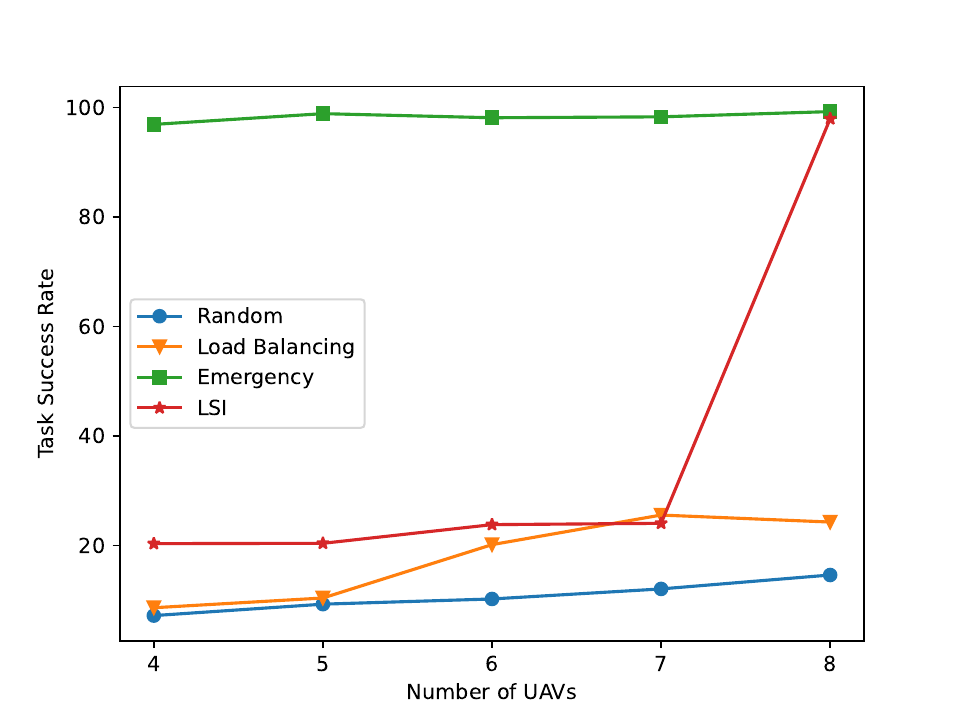}
\caption{Results of Town-1 (Destroyed)} \label{Town1-Res}
\end{subfigure}
\hspace*{\fill} 
\begin{subfigure}{0.32\textwidth}
\includegraphics[width=\linewidth]{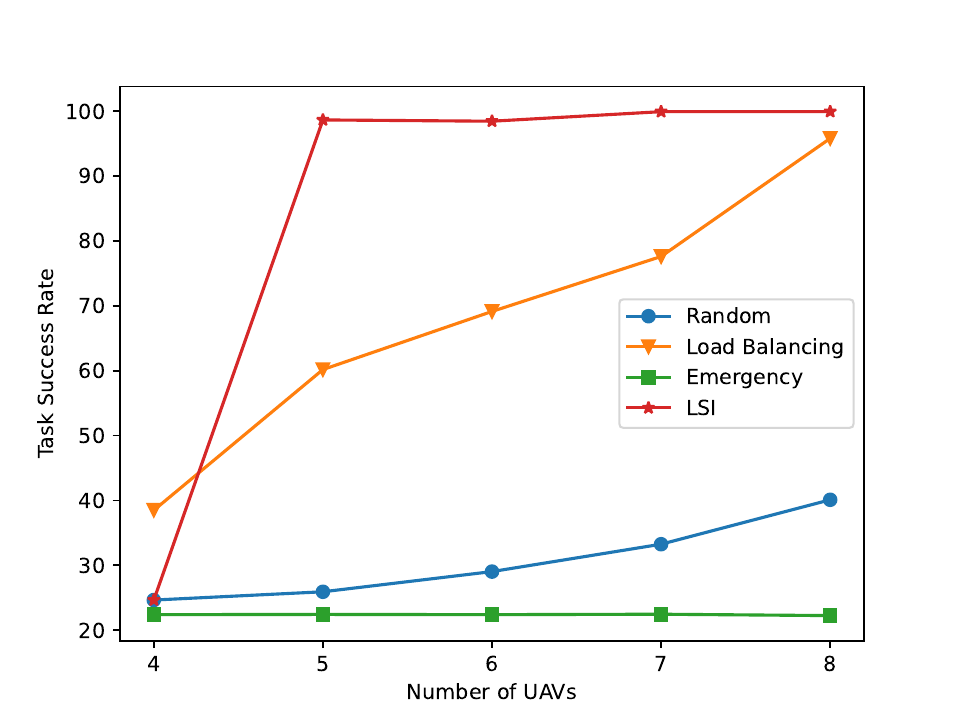}
\caption{Results of Town-2 (Affected)} \label{Town2-Res}
\end{subfigure}
\hspace*{\fill} 
\begin{subfigure}{0.32\textwidth}
\includegraphics[width=\linewidth]{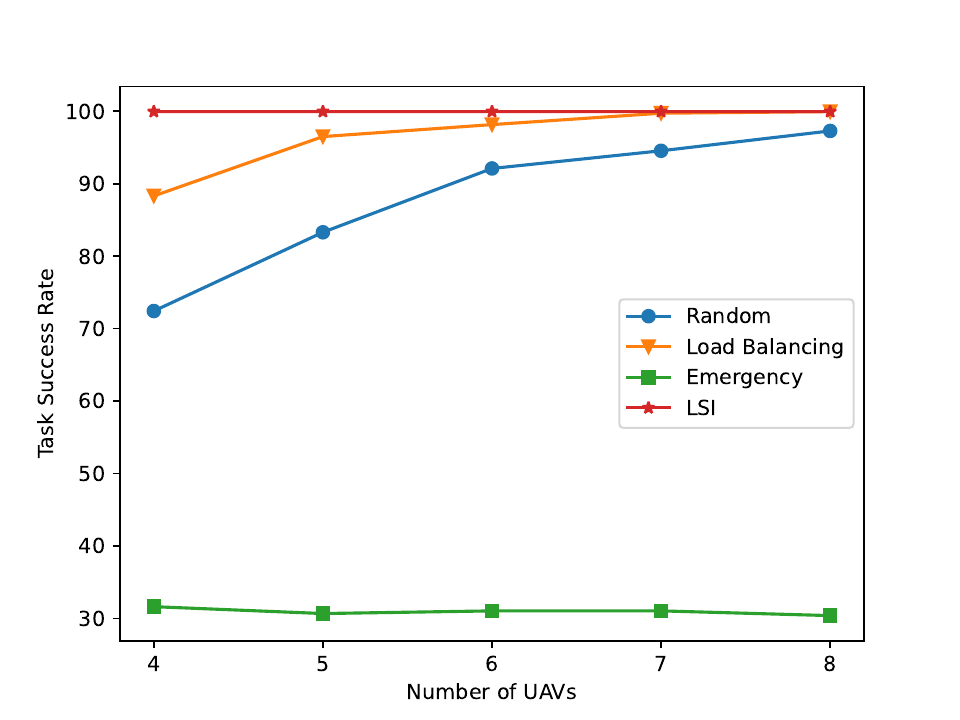}
\caption{Results of Town-3 (Sensed)} \label{Town3-Res}
\end{subfigure}
\caption{The results of each town for different UAV deployment methods.}
\label{TownResults}
\end{figure*}

\begin{figure}[t]
\centering
\includegraphics[scale=0.63]{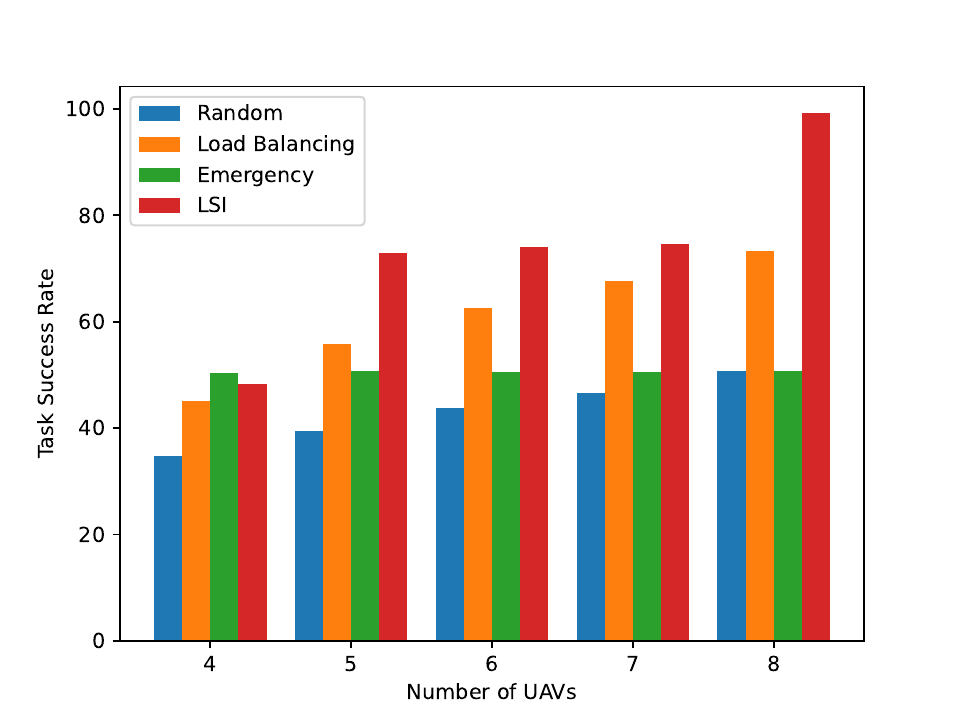}
\DeclareGraphicsExtensions.
\caption{Overall task success rate performance of UAV deployment methods based on the number of UAVs.}
\label{Overall}
\end{figure}

In simulations, we used edge servers as terrestrial resources. Note that we did not consider cloud servers for offloading since majority of the tasks would be failed considering the wide area network (WAN) delay and required worst-case delay of tasks. Moreover, WAN performance could also be affected by the disaster. Each edge server in each town is identical; an edge server has a capacity of 100K CPU units/sec. We assume that each user in three towns is in the range of an edge server so that they can offload their tasks.

Each UAV in the simulation environment is identical in terms of capacity, radius, and altitude. To this end, a UAV has a capacity of 50K CPU units/sec, a horizontal radius of 100 meters for the offloading range, and an altitude of 200 meters. Note that a UAV does not receive an offloaded task when it is flying towards its destination. Therefore, a task can only be offloaded to a UAV when it arrives to its deployed location. Moreover, all of the offloaded tasks in a UAV queue are processed and returned to the corresponding user regardless of UAV flying state. 

In our simulations, we consider only the offloaded tasks. If a user is in the range of both an edge server and a UAV, a task is offloaded to the corresponding resource whose available time is the closest. To this end, we assume that users are informed by the HAP about the queueing conditions of the resources. The fixed WLAN delay used in simulations is 1ms for the tasks offloaded to an edge server, while it is 5ms for the tasks offloaded to a UAV.

\begin{figure*}[!t]
\begin{subfigure}{0.45\textwidth}
\includegraphics[width=\linewidth]{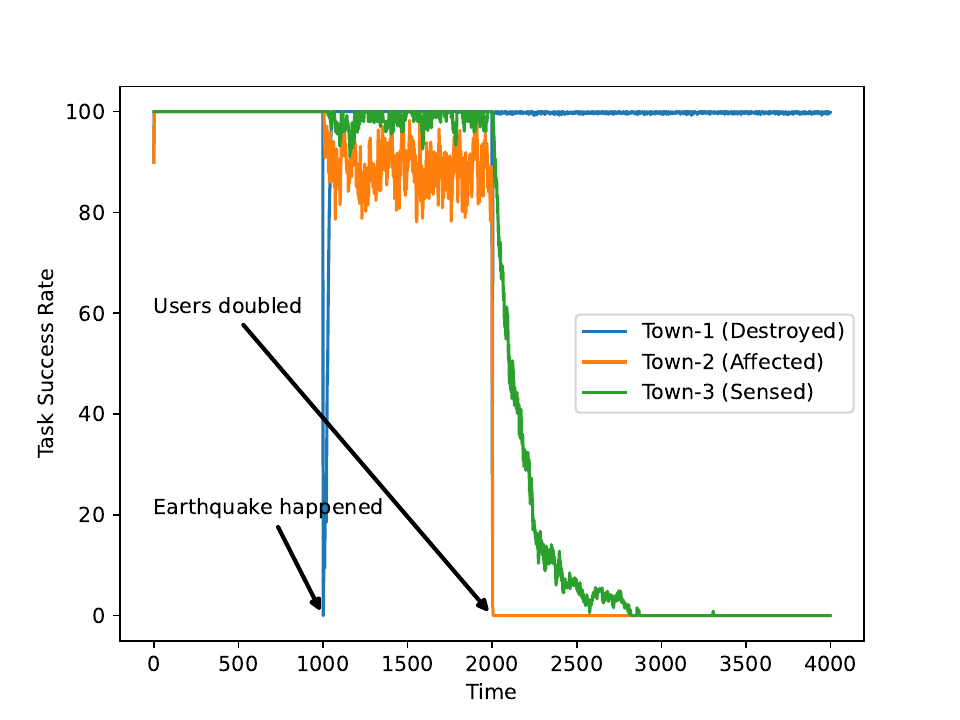}
\caption{The Emergency Method} \label{EmergencyTime}
\end{subfigure}
\hspace*{\fill} 
\begin{subfigure}{0.45\textwidth}
\includegraphics[width=\linewidth]{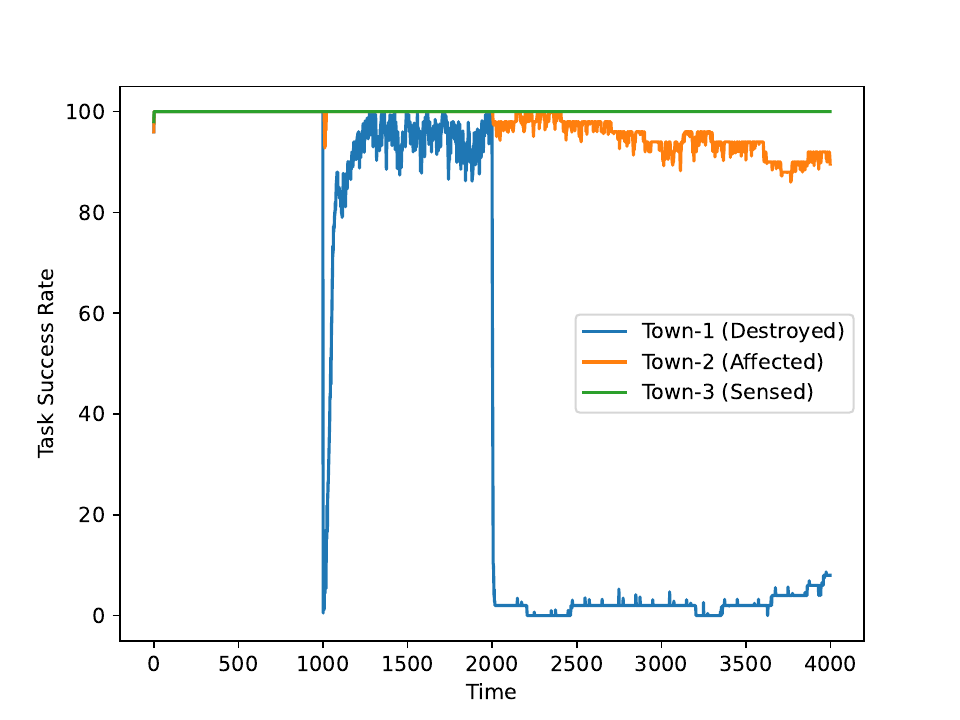}
\caption{The Load Balancing Method} \label{LoadBalancing-Time}
\end{subfigure}
\begin{subfigure}{0.45\textwidth}
\includegraphics[width=\linewidth]{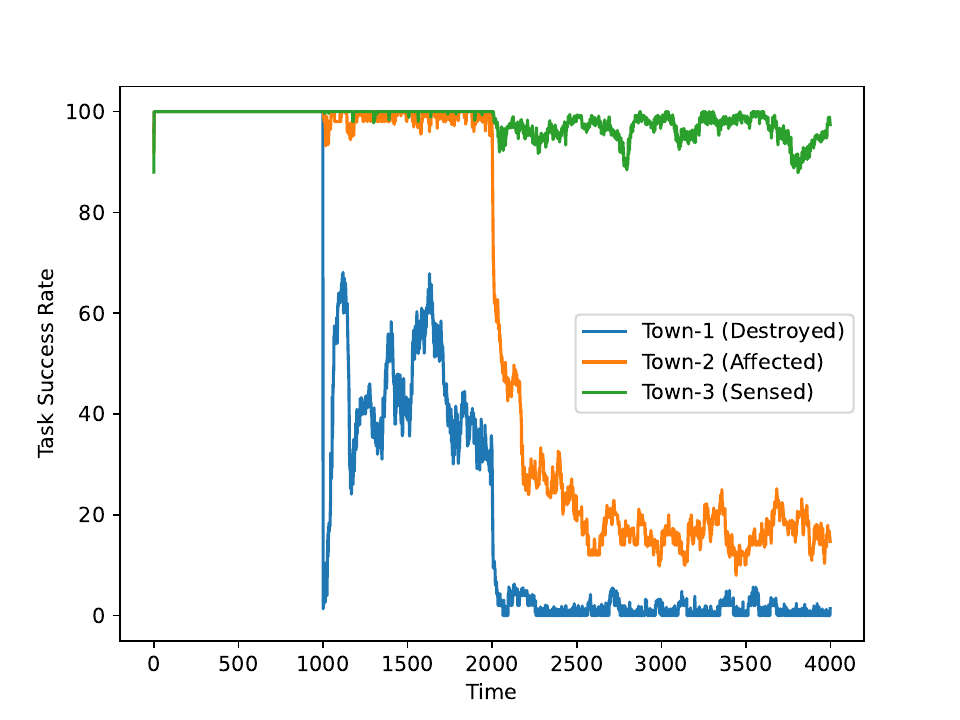}
\caption{The Random Method} \label{Random-Time}
\end{subfigure}
\hspace*{\fill} 
\begin{subfigure}{0.45\textwidth}
\includegraphics[width=\linewidth]{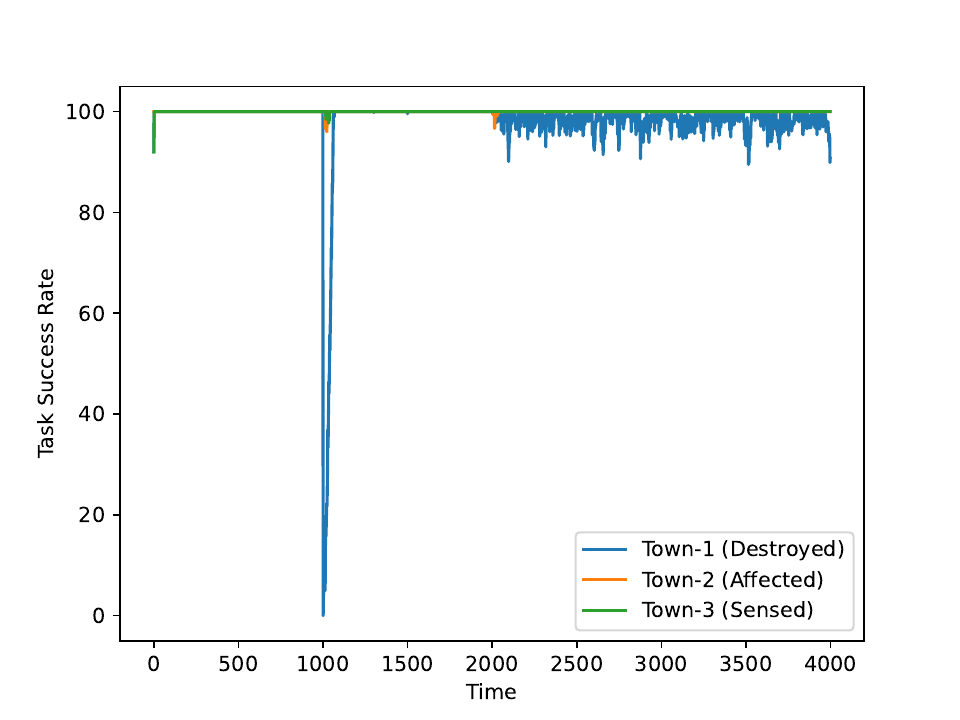}
\caption{The LSI Method} \label{LSI-Time}
\end{subfigure}
\caption{ The effect of earthquake and UAV deployment methods in time considering 8 UAVs.}
\label{OverallTimeResults}
\end{figure*}

\subsection{Results}

We first observed the outcome of the policy of no UAV case, which can be considered as the baseline. As shown in Figure \ref{NoUAVTime}, the task success rate decrease to 0\% after the earthquake at 1000 seconds since the infrastructure, which is the edge server in our scenario, is destroyed. Therefore, the corresponding tasks of the users in Town-1 cannot be processed. Moreover, since no UAV is used after the earthquake, the condition of Town-1 does not improve. On the other hand, the task success rate in Town-2 and Town-3 starts to oscillate after the earthquake since the applications are used more frequently so that the capacity of the existing infrastructure cannot be sufficient completely. Especially, when the number of users doubles after 2000 seconds, the task success rate of Town-2 decreases to 0\% regarding the delay requirements of the tasks. Besides, the decline of the task success rate in Town-3 is not as sharp as in Town-2 since the delay requirements of tasks in Town-3 can provide more tolerance to queueing delay in the edge server. It is important to note that each town produces the same amount of tasks.


Initially, we evaluated the performance of UAV deployment methods for each town. Afterwards, we compared the overall performances. Note that since the results are plausible for each method after 4 UAVs, we assessed the results starting from 4 UAVs. The results for each town are shown in Figure \ref{TownResults}.


Considering Town-1, the performance of the emergency method is superior until 8 UAVs as shown in Figure \ref{Town1-Res}. This is the expected result as all available UAVs are directed to the disaster area when the emergency method is deployed. Since other methods also take the conditions of other towns into account, they send the corresponding UAVs to other locations in addition to the disaster area. The effect of this consideration manifests itself in the results of Town-2 and Town-3 as shown in Figures \ref{Town2-Res} and  \ref{Town3-Res}, respectively. The emergency method is the worst method for those two towns since it does not allocate any UAVs for those locations. On the other hand, the LSI method outperforms other methods since it allocates UAVs based on the load and delay requirements of the tasks in those towns. Since the load balancing method assigns UAVs to towns based on the  number of tasks and does not evaluate the delay requirements, it could not provide as sufficient task success rate as the LSI method.  

Apart from the town-based results, we also evaluated the overall performance of UAV deployment methods. As shown in Figure \ref{Overall}, the LSI method outperforms other approaches. Note that the only exception with four UAVs is originated by the fact that the emergency method sends all UAVs to the disaster area. 

Since we are proposing a mechanism for dynamic capacity enhancement, we also evaluated the performance in time to observe the impact of the earthquake and the efficiency of UAV deployment methods more clearly. To this end, we analyzed the case in which each method uses 8 UAVs. As shown in Figure \ref{OverallTimeResults}, the task success rate of Town-1 drops drastically for each method when the earthquake happens. Afterwards, its recovery varies based on the reactive quality of the corresponding method. For example, as shown in Figure \ref{EmergencyTime}, the recovery in the emergency method is rapid, and it is not  affected by the doubled number of users after 2000 seconds since all of the UAVs are deployed in the disaster area. However, since the dynamic capacity enhancement cannot be carried out for Town-2 and Town-3, their task success rate oscillates and then declines, respectively. The fast recovery also occurs in the load balancing method as shown in Figure \ref{LoadBalancing-Time}. However, since it deploys UAVs evenly considering the number of tasks for each town, the success rate of Town-1 declines rapidly after 2000 seconds. Moreover, the enhanced capacity of Town-2 also causes oscillations in task success rate since it cannot meet the required delay for each task. The random method shown in Figure \ref{Random-Time} causes poor results for each town, especially Town-1, and Town-2. Finally, the LSI method provides the highest task success rate as shown in Figure \ref{LSI-Time}. The main reason of its success is that it takes the delay requirements of applications for each town into account and deploys each UAV to those towns based on their demand.

\section{Conclusion and Future Work}

In this study, we proposed a new paradigm called air computing by extending edge computing to multiple air platforms. Considering the ever-growing delay-intolerant application demands, we believe that air computing would be the next-generation computational paradigm to solve QoS-related issues using 3D networking technologies.

To show an important aspect of the benefits of air computing, we studied an earthquake case by applying efficient UAV deployment methods under HAP management in order to enhance the capacity of the affected areas dynamically. To this end, we applied a scenario in which three towns are affected by the earthquake with different severity. Thus, we applied an emergency method, which considers only the destroyed area, an load balancing method, which deploys UAVs based on the number of tasks produced, a random method, and the LSI method, which computes the delay requirements and required number of UAVs of each town for the UAV deployment.

The experimental results showed that the applied methods should be chosen based on the goals in the environment. For example, if the importance of the task success rate in the disaster area is the most prominent, then the emergency method should be applied. Moreover, if the number of available UAVs is sufficient based on the capacity of those UAVs, the LSI method can also be applied considering other towns. On the other hand, if each town is important regarding different functions of them, then the LSI method should be applied. Note that the load balancing method would also be useful if the task requirements in each town are similar regarding their worst-case delay.

All of the methods in this study are reactive since they deploy UAVs after the disaster has happened. However, considering a dynamic environment in which users are also mobile, a proactive approach in which UAVs can be deployed to the corresponding attraction points could be useful. Therefore, we plan to develop a new proactive dynamic capacity enhancement scheme as future work.

\section*{Acknowledgment}

This work is supported by the Turkish Directorate of Strategy and Budget under the TAM Project number 2007K12-873.

\section*{References}
\bibliographystyle{elsarticle-harv} 

\bibliography{AirComputing}

\end{document}